# GAASP: Genetic Algorithm Based Atomistic Sampling Protocol for High-Entropy Materials

G. Anand[a]*

[a]*Department of Metallurgy and Materials Engineering, Indian Institute of Engineering Science and Technology, Shibpur, Howrah, India*

*Email : ganand@metal.iiests.ac.in

# GAASP: Genetic Algorithm Based Atomistic Sampling Protocol for High-Entropy Materials

High-Entropy Materials are composed of multiple elements on comparatively simpler lattices. Due to the multicomponent nature of such materials, the atomic scale sampling is computationally expensive due to the combinatorial complexity. We propose a genetic algorithm based methodology for sampling such complex chemically-disordered materials. Genetic Algorithm based Atomistic Sampling Protocol (GAASP) variants can generate low and well as high-energy structures. GAASP low-energy variant in conjugation with metropolis criteria avoids the premature convergence as well as ensures the detailed balance condition. GAASP can be employed to generate the low-energy structures for thermodynamic predictions as well as diverse structures can be generated for machine learning applications.

**Introduction**

There has been significant interest in the area of high-entropy and medium-entropy materials due to their interesting properties, such as high-strength and high toughness at liquid helium [1], as well as liquid nitrogen temperature [2], high corrosion resistance [3], strength retention at high temperature [4], high catalytic activity [5], etc. Medium and high-entropy materials containing three or more elements, respectively exhibit these exciting properties, but also present an unique challenge in designing such materials with high computational complexity. The compositional complexity in such materials leads to the combinatorial explosion and traditional hit-and-trial approach for designing such material might not be a feasible approach. Such a problem becomes further challenging with the advent of non-concentrated compositionally complex alloys, where certain element may be added in the dilute amount to the concentrated complex alloy. Dragoe *et. al.* presented a simple calculations to demonstrate the combinatorial explosion through a scenario, in which if 5 elements are chosen from the palette of 26 elements with 1% increment, one can have 2,822,599,802,880 combinations, while of 6 elements are chosen with similar increment in the composition, the possibilities increase to the value of 902,943,619,878,430 [6].

Hence, the composition screening of such materials poses a formidable challenge. Special quasi-random structure (SQS) generation is one of the earliest techniques, which was extensively applied to simulate the chemical disorder in high-entropy materials. But, SQS technique suffers from limited computational efficiency for the multicomponent materials. It involves optimisation of numerous parameters such as cut-off, simulated annealing temperatures, optimisation steps, etc., which makes this technique challenging for optimising HEA structures [7]. Mean-field approach, such as coherent-potential approximation (CPA) has been applied to simulate the chemical disorder [8], however it cannot introduce local short-range order, which influences the properties of high-entropy materials [9]. In view of challenges associated with simulation of high-entropy materials, small set of ordered structures (SSOS) approach was developed for efficient ab-initio calculations for rapid compositional screening of multicomponent materials. In this approach, initially large set of small-ordered structures (SOS) are generated from which small set of ordered structures are chosen. The property of the particular multicomponent material is then calculated from the these small set of ordered structures with appropriate weight assigned to each of SSOS [10]. SSOS has also been extended to the non-equiatomic high entropy materials [11]. Machine learning approaches has also been developed for generating and optimising the composition of high-entropy materials. Neural structure evolution strategy involves application of combined artificial neural network with evolutionary algorithms to generate the large supercell of high entropy alloys with high computational efficiency [12]. Database oriented ANN based machine learning model is employed to predict the structure of HEA was proposed. However, biased databased towards certain family of alloys limits the predictive capability of such approaches [13]. ML-based approaches has also been applied for HEA compositions with high catalytic activities [14]. Initially, experimental approaches were being developed for combinatorial exploration of compositional landscape of high-entropy materials [15]. But, recently there has been marked interest in development of combined experimental-theoretical approaches for compositional screening of high entropy materials. Empirical and experimental approach has been developed for the compositional screening to determine the solid solution and intermetallic phases in high-entropy alloys [16]. Materials informatics based approaches has been employed to screen for compositions with desired properties. Genetic algorithm was used to determine the desired alloy compositions, which were further validated with the experimental fabrication of such alloys [17]. Evolutionary algorithm, such as genetic algorithm (GA) has been applied in some of the above-stated studies. However, GA has been already employed for crystal structure predictions in USPEX [18] for range of materials. GA based codes such as GAtor has been developed molecular crystals [19].

The high-entropy materials represents the unique challenge due to its compositional as well as combinatorial complexities. The compositional complexities has been tacked using GA in the earlier investigations.

GA has been extensively applied to study the compositional disorder. Initial investigation of GA for determining low-energy structures in binary alloys showed its computational efficiency in comparison of Monte Carlo method [20]. GA has been applied to optimise the cation ordering in oxides [21], structure prediction in disordered alloys [22], for studying non-ideality in mixing behaviour of mineral solid solutions [23]. GA suffers from the premature convergence issue and it has been modified to include symmetry adapted crossover [24]. Adaptive GA has been developed to combine classical interatomic potential and Density Functional Theory (DFT) calculations to predict structures of crystal, surfaces and interfaces with higher computational efficiency in comparison of purely DFT based approaches [25]. Atomistic sampling of high entropy materials poses significant challenges in comparison of traditional materials. Firstly, the combinatorial complexity of such materials is significantly higher in comparison of traditional materials, which are generally binary mixtures. Higher compositional complexity warrants the need of high-throughput approach for compositional screening. In view of the above, we propose genetic algorithm based sampling involving alchemical swaps and classical interatomic potential for generation of low-energy structures. Note that present approach can be extended to DFT based approaches as well. We would additionally demonstrate the GAASP can also be employed for generating high-energy structures, which may be relevant for generating database for machine learning applications, where the diversity of configurations may be required for training machine learning model. We would also present the applicability of GAASP for compositional screening.

**Materials and Methods**

*Method*

The aim of the GAASP is to generate the new configurations with desired energy distribution. Such an energy distribution might be biased towards low energy configurations or it might be high energy configurations. The aim of generating low-energy configurations might be for thermodynamic predictions, while high energy distributions might be required as representative configurations for holistic database for machine learning applications. The GAASP approach should be able to generate both of these extreme cases, which would be shown later in sections. The GAASP can be principally divided into the following steps:

*Random sampling to generate the parent population*

Initially, random sampling of atomistic configurations needs to be carried out to sample the potential energy landscape of the system of interest. It is of the paramount importance that truly random sampling is carried out to sample the representative configurations from throughout the potential energy landscape. The number of configurations sampled randomly signifies the '*thoroughness*' of the random sampling (say, this number is N).

*Parent selection*

Once the random sampling is complete the parent population is chosen from the random samples using Roulette wheel selection mechanism (Fig. 1 (a)). The parameters for the Roulette wheel parent selection are number of parents chosen from the original parent population (*i.e.,* N) or X, number of intervals in which the wheel is being divided (n) and number of parents in each interval $\left(A = \frac{N}{n}\right)$ and dividing factor for each interval or D (i.e., first interval is divided into $\frac{1}{2}$, later into $\frac{1}{4}$ and so on). A random number, x is initially generated, such that $0 \leq x < D^n$. The identity of the interval from which a parent is being chosen is dependent whether ,

$$x \in \begin{cases} \left[\frac{D^{b-1}-1}{D^{b-2}}, \frac{D^b-1}{D^b}\right) & If, b \text{ is not the last interval} \\ \left[\frac{D^{b-1}-1}{D^{b-2}}, D^b\right) & If, b \text{ is the last interval} \end{cases}$$

Once the interval is determined, the parent from the particular interval is chosen by choosing a random number between $[0, A)$.

*Information encoding*

Once the parents population is chosen depending upon the desired energy distribution, the parents are chosen sequentially in pair to generate two children. The encoding of the atomic configurations is carried out as value encoding, with values signifying the atomic type as well as the cartesian coordinates of the particular atom. Such encoded information represents the chromosome and the crossover between two chromosomes leads to the generation of newer configurations (Fig. 1 (b)). However in the encoding procedure, it is ensured that certain types of atoms are encoded together, *i.e.,* all the type-1 atoms are encoded together and type-2 after that and so on.

*Alchemical swap process*

The aim of the swapping process is to interchange the information concerning the position of atoms between two configurations, as shown in Fig. 2 (a). Initially, a random number in the range $[0, P]$ is generated, where P is the total number of atoms in the supercell. If we get a number 'i' in the process, it would correspond to same type of atom, but with different cartesian coordinates ($\{X_A, Y_A, Z_A\}$ and $\{X_B, Y_B, Z_B\}$ in parent-1 and parent-2, respectively) in both the supercells due to the type of value encoding employed in the present method (Fig. 2 (b)). The cartesian coordinates of $i^{th}$ atom in parent-1(i.e. $\{X_A, Y_A, Z_A\}$) is searched in parent-2 and say, $k^{th}$ atom in the parent-2 has same cartesian coordinates. Similarly, the coordinates of $i^{th}$ atom in the parent-2 ($\{X_B, Y_B, Z_B\}$) is searched in parent-1 and say $j^{th}$ atom in parent-1 has same coordinates. Once the identity of '$i$' and '$j$' in parent-1 and '$i$' and '$k$' in parent-2 is established, the alchemical swap between '$i$' and '$j$' in parent-1 and swap between '$i$' and '$k$' in parent-2 is carried out (Fig. 2(d) and (e)). Such a swap leads to the occupation of atom in children with inheritance from both the parents. But it also leads to the mutation, involving the introduction of different atoms at the original position ('$i^{th}$ position' in parent-1 and parent-2). Such a method ensures that the composition of the supercell remains constant during the swap process.

The important parameters associated with the swap process is number of swap (Q), such a number is taken randomly between 10-30% of the total number of atoms in the supercell (P).

*Energy calculation*

Once the parent population for a particular genetic algorithm cycle is generated, energy calculations in the NVT ensemble is carried out for all the parents by running molecular dynamics simulation for 10000 steps with first 1000 steps are solely employed for equilibration. In the present investigation, we carry out the energy calculations for BCC-AlCoCrFeNi HEA as proof-of-concept studies. The GAASP approach can be coupled with DFT-based codes for the energy calculations, in principle. We employed the EAM potential developed by Zhou et. al. [26] for Al, Co, Fe and Ni, while for Cr modified Zhou-EAM potential [27] was employed to describe AlCoCrFeNi. Energy for each of the configuration or parent is stored. After energy calculations, the energy values are sorted in the increasing order and configurations are arranged in the similar manner. The parents for the next generation are again chosen from this reservoir using the Roulette wheel selection mechanism.

**Results and Discussion**

*Generation of low-energy and high-energy configurations*

Figure 3(a) shows the ridge plot of the evolution of the configurational energy distribution as the GA cycle progresses. It can be seen that GAASP generate the lower energy configurations. Such low-energy landscape can be relevant for the prediction of thermodynamic properties of such materials, which are dictated by low-energy configurations. Similarly, Fig. 3(b) shows the evolution of high energy states in high-energy variant of GAASP. The high-energy or non-equilibrium structures are often required in efficient sampling strategies [29]. The high energy variant of GAASP choses the parent in the reverse order in comparison to low-energy variants as discussed in the method section. Note that the applicability of high-energy GAASP variant in conjugation with low-energy variant can be used in generation of the diverse configurations for training ML models, which not only requires equilibrium, but non-equilibrium structures in training.

*Comparison of GAASP and GAASP-metropolis method for generation of the genetic diversity*

We have additionally extended the GAASP approach to include the metropolis acceptance criteria (GAASP-metropolis) to ensure the detailed balance in the sampling [30]. In such an acceptance criteria, the child is accepted on the basis of energy change. If, $\Delta E$ or change in energy due to atomic swap to generate a child from parent is less than or equal to the zero, then child is accepted, while is it is greater than zero, than child is accepted with the probability, $P = e^{\left(\frac{-\Delta E}{k_B T}\right)}$. Application of the metropolis criteria for child acceptance serves two purposes. First, it ensures the detailed balance in the sampling and secondly, it helps in avoiding the premature convergence of GA sampling. The existence of the detailed balance in sampling ensures that the energy landscape explored through GAASP approach can be employed for determining the thermodynamic properties of material being simulated. The premature convergence of genetic algorithm based sampling is well-known issue [22] and in the present investigation, we demonstrate that employing the metropolis criteria can avoid such an issue, while it can also lead to better exploration of the energy landscape. Figure 4 shows the GAASP without metropolis acceptance criteria reaches convergence faster, but GAASP with metropolis criteria generates lower-energy structures and it is better in exploring the composition landscape of AlCoCrFeNi HEA by generating configurations with lower configurational energies.

*GAASP with compositional variation*

We have additionally extended the GAASP to modify the compositions of the configurations to either generate high or low energy structures. In such an approach, the alchemical swap between two configurations are carried out without mutation. Figure 5(a) shows the change in the configuration energy of the supercell with GA cycle for both energy increase and decrease objective of GAASP. Note that in the context of constant composition, with each swap between parents there is associated mutation. In the present scenario, we have simply carried out the swap without any explicit mutation. However, random mutation can be introduced, in principle and effect of the mutation would be included in the future studies. Figure 5 (b) shows the change in the number of elements as GAASP explores the low-energy landscape. It can be seen that Fe is increasing in the supercell, while Co and Ni show marginal increase, while Cr and Al show decreasing trend. Figure 5(c) shows the variation in the number of elements in the AlCoCrFeNi supercell, which shows the number of Al atoms increase as the higher-energy , while other elements show the marginally decreasing trend. It should be noted that we are demonstrating the applicability of GAASP for simulating the composition variation as either function of increasing or decreasing energy as objective and corresponding change in the composition can be predicted for high entropy materials. Such an approach can be particularly important for exploring non-equiatomic compositions of high-entropy materials.

**Conclusions**

In the present investigation, we have developed the Genetic Algorithm based Atomistic Sampling Protocol (GAASP) for efficient sampling of atomistic structure of compositionally disordered multicomponent materials or high-entropy materials through alchemical swap process. GAASP approach can be employed to generate the low-energy as well as high energy structures. GAASP approach can be used to generate the diversity of configurations, which may be relevant for generating diverse configurations for training machine learning models for high-entropy materials. GAASP approach can also be applied to for optimisation of composition of such multicomponent materials for determining the compositions, which can lead to high-energy or low-energy configurations.

**Code**

The GAASP code is available at: https://github.com/ganand1990/GAASP

**Acknowledgements**

Author is thankful to Dr. Colin Freeman for suggesting to develop the protocol and Prof. Graeme Ackland for the critical feedback. We are also thankful to National Supercomputing Facility for access to Param Sidhi AI system.

**Figures**

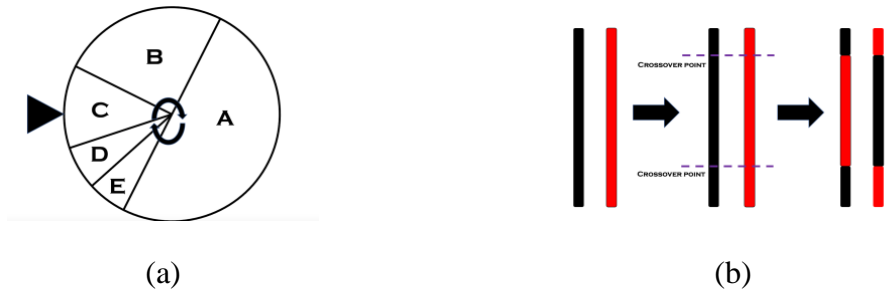

(a)                                                 (b)

Figure – 1: (a) Schematic of the Roulette wheel selection process and (b) genetic crossover of information to generate new configurations.

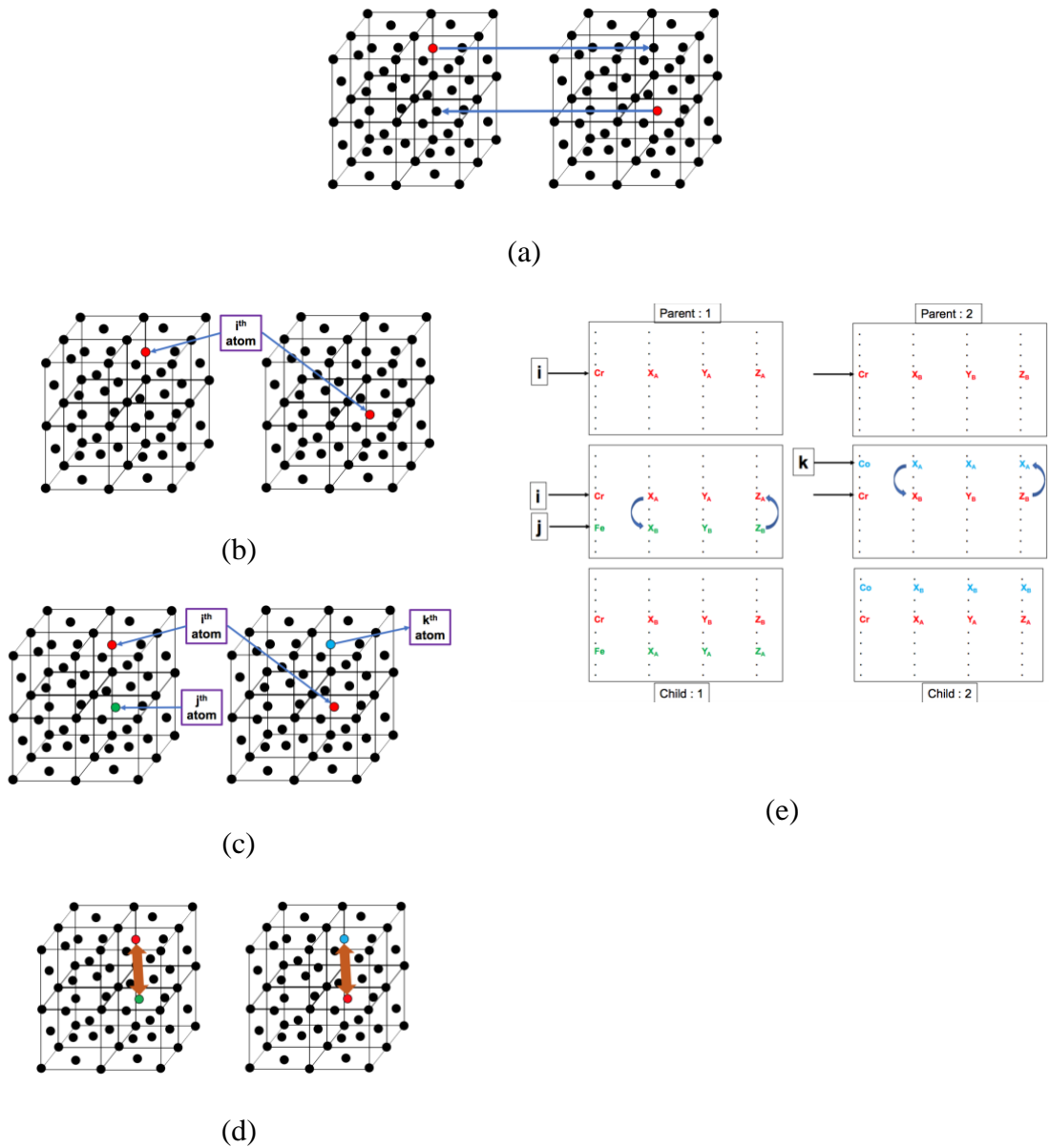

Figure-2: (a) Aim of the alchemical swap, *i.e.,* swap the information about the existence of atoms at particular lattice site in the supercell, (b) The value encoding is carried out in a way to ensure that same atom-type is present at $i^{th}$ index., (c) process of determination of $j^{th}$ atom in parent-1 with same coordinates as $i^{th}$ atom in parent-2 and $k^{th}$ atom in parent-2 having same coordinates as $i^{th}$ atom in parent-1, (d) swap between $i^{th}$ and $j^{th}$ atoms in parent-1 and $k^{th}$ and $i^{th}$ atoms in parent-2 and (e) demonstration of steps (b-d) as implemented using value-encoding in GAASP code [28].

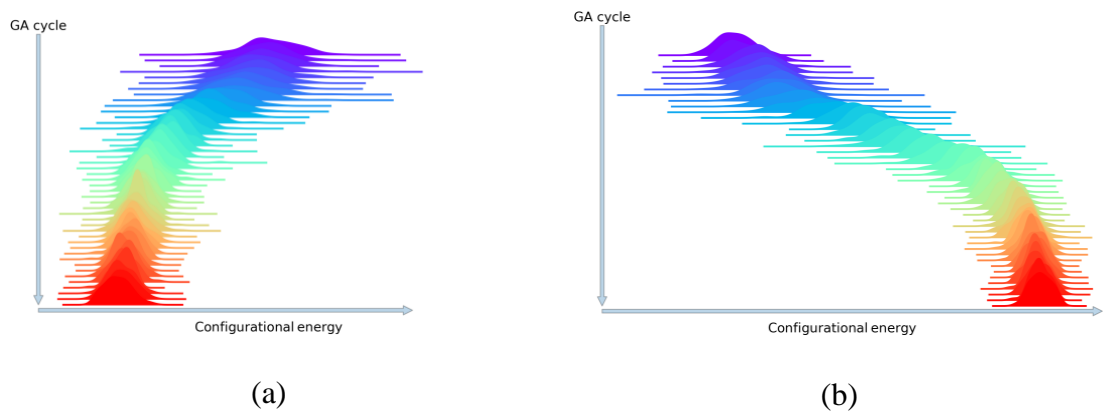

(a)                           (b)

Figure-3: Ridge plots of the energy distribution as obtained from the (a) low-energy and (b) high energy search for AlCoCrFeNi HEA.

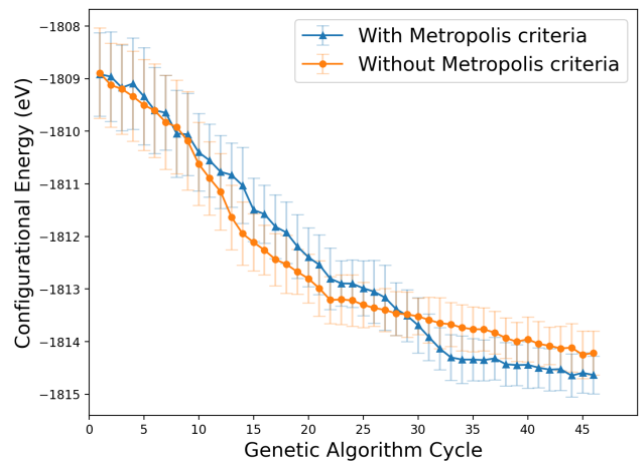

Figure-4: Comparison of variants for GAASP for generating low-energy structures involving metropolis criteria for selection of children and no rejection (i.e. without metropolis criteria).

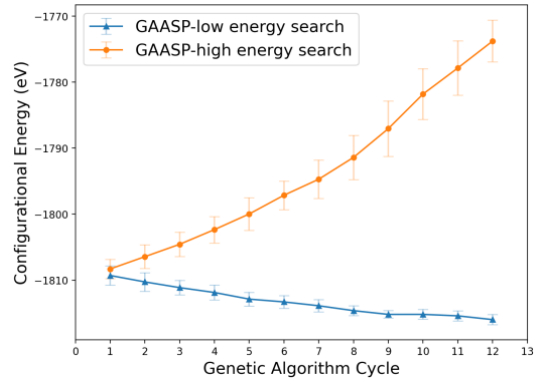

(a)

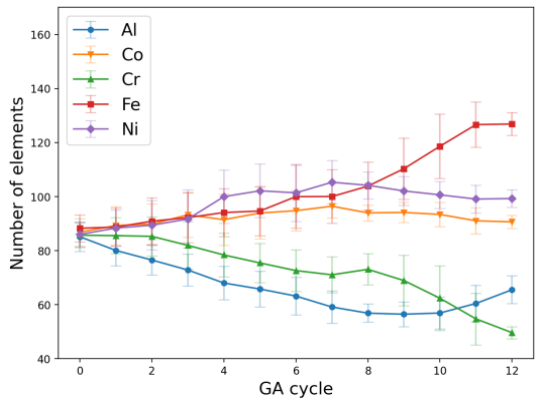

(b)

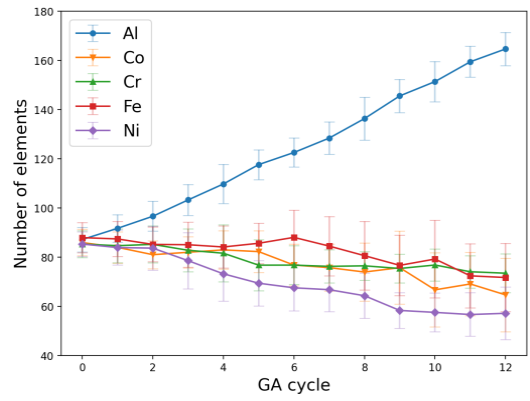

(c)

Figure-5 : (a) Change in the energy for composition variation in the supercell in GAASP for low and high energy search, (b) change in the number of atoms for GAASP low-energy search and (c) high-energy search.